\documentclass[conference]{IEEEtran}
\IEEEoverridecommandlockouts
\usepackage{cite}
\usepackage{amsmath,amssymb,amsfonts}
\usepackage{algorithmic}
\usepackage{graphicx}
\usepackage{textcomp}
\usepackage[table]{xcolor}
\definecolor{mygreen}{HTML}{99ffff}
\definecolor{myblue}{HTML}{ccffcc}
\def\BibTeX{{\rm B\kern-.05em{\sc i\kern-.025em b}\kern-.08em
    T\kern-.1667em\lower.7ex\hbox{E}\kern-.125emX}}
\usepackage{array}
\usepackage{booktabs}
\usepackage{multirow}
\usepackage{subfig}
\usepackage{arydshln}
\usepackage{url}

\begin{document}

\title{A Multi-Label EEG Dataset for Mental Attention State Classification in Online Learning \\
\thanks{\textsuperscript{\#}Equal contribution, \textsuperscript{*}Corresponding author.}
}

\author{
Huan Liu\textsuperscript{1,2,\#}, Yuzhe Zhang\textsuperscript{1,\#,*}, Guanjian Liu\textsuperscript{1}, Xinxin Du\textsuperscript{3}, Haochong Wang\textsuperscript{4}, Dalin Zhang\textsuperscript{5} \\
\textit{School of Computer Science and Technology, Xi'an Jiaotong University, Xi'an, China\textsuperscript{1}} \\
\textit{BigKE Joint Innovation Center, SHMEEA, Shanghai, China\textsuperscript{2}}\\
\textit{Department of Computer Science and Technology, Tsinghua University, Beijing, China\textsuperscript{3}} \\
\textit{School of Life Sciences and Technology, Xi'an Jiaotong University, Xi'an, China\textsuperscript{4}} \\
\textit{Department of Computer Science, Aalborg University, Aalborg, Denmark\textsuperscript{5}} \\

}


\maketitle

\begin{abstract}
Attention is a vital cognitive process in the learning and memory environment, particularly in the context of online learning. 
Traditional methods for classifying attention states of online learners based on behavioral signals are prone to distortion, leading to increased interest in using electroencephalography (EEG) signals for authentic and accurate assessment. 
However, the field of attention state classification based on EEG signals in online learning faces challenges, including the scarcity of publicly available datasets, the lack of standardized data collection paradigms, and the requirement to consider the interplay between attention and other psychological states. 
In light of this, we present the Multi-label EEG dataset for classifying Mental Attention states (MEMA) in online learning. 
We meticulously designed a reliable and standard experimental paradigm with three attention states: neutral, relaxing, and concentrating, considering human physiological and psychological characteristics. This paradigm collected EEG signals from 20 subjects, each participating in 12 trials, resulting in 1,060 minutes of data. Emotional state labels, basic personal information, and personality traits were also collected to investigate the relationship between attention and other psychological states. Extensive quantitative and qualitative analysis, including a multi-label correlation study, validated the quality of the EEG attention data.
The MEMA dataset and analysis provide valuable insights for advancing research on attention in online learning. The dataset is publicly available at \url{https://github.com/XJTU-EEG/MEMA}.
\end{abstract}


\section{Introduction}
Attention refers to the ability to focus on task-related information stimuli for a relatively long period of time while consciously attempting to ignore other stimuli~\cite{fortenbaugh2017recent}. It is essentially governed by three networks in the brain: alertness, orientation, and execution~\cite{petersen2012attention}. Attention is closely related to a learner's education quality as a crucial cognitive process in the learning and memory environment. With the rise of online learning, the online learning environment, characterized by a massive number of learners online simultaneously and a lack of supervisors, contrasts sharply with traditional offline learning scenarios. This has made the need for automated assessment of learners' states more urgent and has also drawn increased research focus to this area.

Traditional methods for classifying the attention states of online learners are based on behavioral signals observed in online videos, including speech, facial expressions, gestures, and body movements~\cite{kruger1998attentive, dolcos2022impact}. Although these methods can somewhat reflect the learners' attention states, they are easily influenced by their subjective actions, allowing for the possibility of deliberately deceiving the classification algorithms through active disguises. On the contrary, the classification of online learners' attention states based on electroencephalography (EEG), a physiological signal, has garnered widespread attention due to its authenticity. Meanwhile, EEG has a high temporal resolution by directly measuring brain activity and has been proven to accurately reflect changes in attention states~\cite{du2020efficient}. 
With the increasing portability and affordability of EEG devices, the prospects of employing EEG to monitor learners' attention in online learning scenarios are becoming increasingly promising.



Despite some progress, the field of EEG-based attention state classification in online learning still faces several challenges.
Firstly, a significant drawback is the scarcity of publicly available datasets, and those that are available often suffer from inadequate data quality validation. This deficiency severely hampers the reproducibility and comparability of research in the field~\cite{gupta2021attention, ghasemy2019detection}.
Secondly, the lack of standardized or uniform data collection paradigms and arbitrary processes further compounds the challenge. Various stages lack the incorporation of pertinent psychological and physiological knowledge, such as the selection of collection times, the design of collection processes, and the formulation of experimental tasks~\cite{wang2021eeg, alirezaei2017detection}. Consequently, adhering to a consistent collection paradigm becomes difficult.
Thirdly, research about learners' attention is often overly simplistic and idealized, leading to limited data, labels, and gathered information. It is worth noting that attention is closely intertwined with other psychological states, such as emotional states~\cite{hu2016attention, liu2024libeer}.

The above discussion indicates that this field is still hungry for fundamental data construction research.
In light of this, we propose a Multi-label EEG dataset for classifying Mental Attention states (MEMA) in the context of online learning.
Specifically, we have meticulously designed a reliable paradigm with three distinct attention states: neutral, relaxing, and concentrating. The details of the paradigm, such as task and duration for each state, have been carefully established, considering human physiological and psychological characteristics to ensure standardization and rationality throughout. Employing this paradigm, we gather EEG signals from a cohort of 20 subjects, each participating in 12 trials, resulting in a total of 1,060 minutes of data.
Additionally, to investigate the relationship between attention and other psychological states, we collect emotional state labels~\cite{zhang2024cross, wu2023}, basic personal information, and subjects' personality traits.
Furthermore, we conducted extensive quantitative and qualitative analysis involving a multi-label correlation study to validate the quality of the EEG attention data.

The contributions of this study are as follows:
\begin{itemize}
	\vspace{-0.05cm}
    \item We have made a fundamental research achievement by constructing a multi-label EEG dataset for classifying mental attention states in the context of online learning.
    \item We meticulously designed a reliable and standardized paradigm tailored for attention, integrating physiological and psychological principles, acquiring EEG signals from 20 subjects for more than 1,000 minutes.
    \item We validated the quality of the EEG attention data through extensive quantitative and qualitative analysis, including a multi-label correlation study. The baseline algorithms achieved accuracy rates of up to 85.12\% and 64.84\% for subject-dependent and cross-subject attention state classification tasks, respectively. 
\end{itemize}

\begin{figure}[!t]
	\centering
	\includegraphics[width=\linewidth]{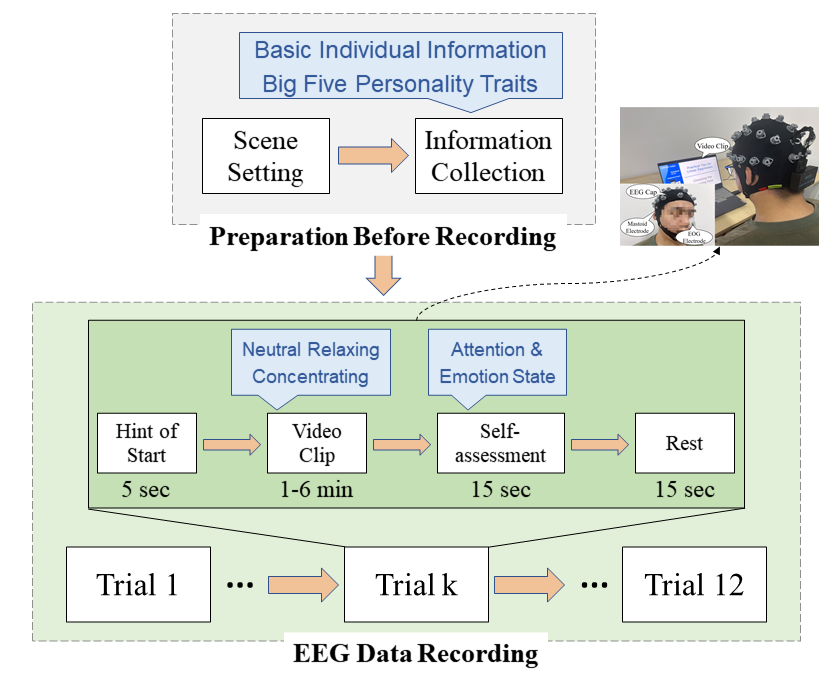}
	\vspace{-0.6cm}
	\caption{Overall collection procedure consists of two parts: preparation before recording and EEG data recording.}
	\label{protocol}
	\vspace{-0.65cm}
\end{figure}
\section{Data Acquisition Paradigm}
This section introduces the data acquisition paradigm, including the collection procedure, task design, and multi-label information collection.

\subsection{Overall Collection Procedure}
The acquisition process of attention EEG signals is depicted in Figure~\ref{protocol}, with each module explained in the following sections. All sessions follow a sequential procedure.

In the ``preparation before recording" phase, the collector sets up the scene, including questionnaires and equipment. Subjects then complete personal information and Big Five Inventory (BFI) questionnaires. Researchers ensure subjects fully understand the data collection process, with details on scene setup and information collection provided later.

During the ``EEG data recording" phase, subjects complete 12 tasks across three types. Each task starts with a 5s hint, followed by a video clip (1 to 6 minutes). Afterward, subjects have 15s for self-assessment of attention and emotional states, followed by a 15s rest, which can be extended if needed. This cycle repeats until all tasks are finished.

\subsection{Preparation before Recording}
\subsubsection{Scene setting}
Data collection occurs in a controlled, well-lit, quiet, and isolated room with essential equipment. The setup includes computers for stimuli delivery and data collection, along with a wireless multichannel EEG system (ZhenTec-NT1-32). The EEG cap, fitted with 32 electrodes following the international 10-10 system, positions the reference electrode at CPz and the ground electrode at FPz.  Electrodes maintain an impedance below 20 $k\Omega$ and sample at 500 Hz, with signals amplified and transmitted by an EEG acquisition amplifier and data receiver. The upper right corner of Figure~\ref{protocol} shows the preparation process.

\subsubsection{Individual information collection}
To enrich the dataset, auxiliary information such as subjects' basic individual details and Big Five personality assessments is gathered alongside EEG signals and corresponding attention and emotion state labels. This data helps explore complex patterns and tasks.

\noindent\textbf{Basic individual information.}
EEG data varies significantly across subjects~\cite{li2018exploring}, closely linked to their individual attributes~\cite{wiberg2019handedness}. Attention levels also correlate with these attributes. Therefore, subjects complete a basic questionnaire covering age, gender, handedness, education level, and current emotional state at the experiment's start.

\begin{table*}[t]
	\caption{Dataset statistical comparison between existing attention EEG datasets and ours.}
	\label{dataset}
	\renewcommand{\arraystretch}{1}
	\centering
 {\small{
            \begin{tabular}{m{3cm}<{\raggedright}m{1cm}<{\centering}m{2.5cm}<{\centering}m{1cm}<{\centering}m{0.8cm}<{\centering}m{2cm}<{\centering}m{2cm}<{\centering}m{2cm}<{\centering}}
                \noalign{\global\arrayrulewidth=2pt}
                \toprule
                 \textbf{Dataset}&\textbf{Public}&\textbf{Scene}&\textbf{\#Subject}&\textbf{\#Trial}&\textbf{Duration}&\textbf{Validation}&\textbf{Extra Data}\\
                \noalign{\global\arrayrulewidth=0.5pt}			
                \midrule
                
                Toa et al. ~\cite{toa2021electroencephalogram}&No&Unclear &30&40&2,400 min&ERP, C&Eye\\
                Alirezaei et al. ~\cite{alirezaei2017detection}&No&mental arithmetic&12&—&273 min&C&None\\
                Ghasemy et al. ~\cite{ghasemy2019detection}&No&auditory attention&10&8&560 min&S, C&None\\
                Gupta et al.
                ~\cite{gupta2021attention}&No&online learning&80&1&800 min&T, S&None\\
                Sethi et al.
                ~\cite{sethi2018eeg}&No&online learning&42&2&—&S&None\\
                Wang et al. ~\cite{wang2021eeg}&Yes&Unclear&1&80&—&S, C&None\\
                Ni et al. ~\cite{ni2020eeg}&Yes&online learning&28&3&420 min&S&I\\
            MEMA(Ours)&Yes& online learning&20&12&1,060 min&B, T, S, C, M &E, I, P\\
            \noalign{\global\arrayrulewidth=1pt}
            \bottomrule
            \end{tabular}
           \\ ERP: event-related potential, C: classification, Eye: eye gaze, S: statistical analysis, T: time-frequency analysis, B: brain electrical activity mapping,   M: multi-label correlation study, E: emotion labels, I: individual information, P: personality 
            }}
            \vspace{-0.5cm}
\end{table*}

%
%
%
\noindent\textbf{Big Five personality assessment.}
One's attention level is influenced by their personality. While various methods measure personality, the Big Five traits are the most widely recognized in scientific research~\cite{zhao2017emotion}. Hence, at the experiment's start, subjects complete the 44-item BFI questionnaire, selecting options that best match their actual conditions.


	
 
   
   

\subsection{Task Design for Each Trail}
Considering individual physiological and psychological characteristics, and previous related dataset-building research~\cite{alirezaei2017detection, ghasemy2019detection}, we categorize attention states into three types: \textit{neutral}, \textit{relaxing}, and \textit{concentrating}, and design specific experimental tasks for each state.

Each subject undergoes four rounds of experiments, each including one task of each type, totaling 12 trials. The task order within each round is randomized to avoid sequence effects that could bias subjects and affect data reliability. The tasks are as follows:

 \textbf{Neutral Task}:  A one-minute blank video is played, and subjects are instructed to maintain a normal state while watching the screen. The ``Neutral" state lies between ``relaxing" and ``concentrating," with subjects neither actively relaxing nor consciously focusing.
 
\textbf{Relaxing Task}: A five-minute soothing video of beautiful scenery, accompanied by relaxing music, is played to help subjects relax and reduce focus, as relaxing music has been shown to aid in physiological and psychological relaxation~\cite{hancock1989dynamic}. Subjects are instructed to stay as relaxed as possible.

\textbf{Concentrating Task}: A five-minute video clip from a well-known machine learning
 course\footnote{https://www.coursera.org/specializations/machine-learning-introduction} is played, covering a knowledge point. This duration is based on research from~\cite{nuechterlein1983visual} on the Continuous Performance Test (CPT) task. Subjects are required to concentrate, actively learn, and think about the material. After the video, subjects answer a simple multiple-choice question related to the content to assess their focus.


After watching the video, subjects perform two mental state assessments: self-assessment of attention and emotion. For the attention self-assessment, they select the option that best matches their attention state from three states. For the emotion~\cite{zhang2022eeg, lou2024, wadlinger2011fixing}, subjects rate their emotions using the ``Valence-Arousal-Dominance" (VAD) model~\cite{schlosberg1954three}, scoring each dimension on a scale from 1 to 5.

\section{Technical Validation}
In this section, we present the constructed multi-label EEG dataset MEMA and conduct comprehensive qualitative and quantitative analysis experiments for quality validation.

\subsection{Dataset Description and Comparison}
As shown in Table~\ref{dataset}, we summarize our collected dataset and several related datasets to facilitate a comparison from various dimensions, including whether it is publicly available, the application scene, the number of subjects sampled, the number of trials per subject, the total duration of EEG data collected, which kinds of data quality validation have been conducted, and which kinds of data beyond attention states have been collected. The table shows that our dataset is quite extensive, only surpassed by one dataset that is not publicly available. We also conduct quality validation to ensure the integrity of our dataset, a feature lacking in other datasets. Moreover, we collect additional data on emotional labels and subjects' personality traits, enabling more complex explorations of cognitive mechanisms on our dataset.

\subsection{Qualitative Analysis}
\begin{figure}[!t]
	\includegraphics[width=\columnwidth]{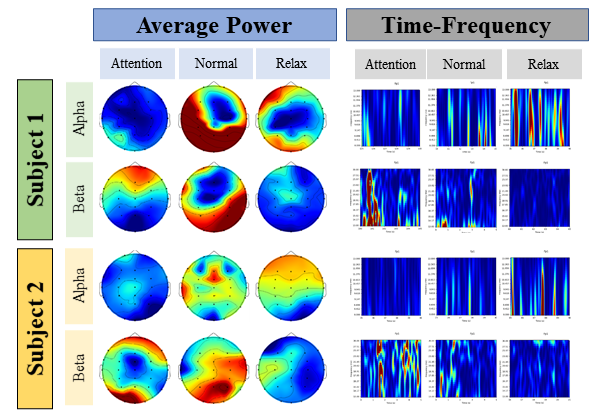}
 \vspace{-0.2cm}
	\caption{Average power results (on topographical maps) and time-frequency of one subject.}
		\label{qualitative}
  \vspace{-0.5cm}
\end{figure}

To reduce noise in EEG signals~\cite{wadlinger2011fixing}, we apply standard preprocessing: a 50 Hz notch filter for power line noise removal, a band-pass filter (8-30 Hz) for relevant frequencies, and Independent Component Analysis (ICA) for artifact elimination. We then qualitatively analyze one subject's brain activity via brain mapping and time-frequency analysis to study activation differences (Figure~\ref{qualitative}). By averaging power in the $\alpha$ (8-13 Hz) and $\beta$ (13-30 Hz) bands and plotting topographic maps focused on the frontal lobe~\cite{duann2009detecting}, we observe that as attention decreases, $\alpha$ power increases while $\beta$ power decreases, consistent with prior studies~\cite{clarke2020resting, magosso2019eeg, misselhorn2019frontal}. This pattern is further validated through time-frequency analysis with the Morlet wavelet transform, showing similar trends in the $\alpha$ and $\beta$ bands at FP1 across three attention states.

\begin{table*}[t]
	\caption{Mean accuracy (ACC, \%) and mean F1 score (\%) of classical methods on the attention state classification tasks  (green areas) and emotion classification (V for valence, A for arousal and D for dominance) tasks (blue areas).}
	\label{Emotion}
	\renewcommand{\arraystretch}{1}
	\centering
	{\small{
            \begin{tabular}{m{1.3cm}<{\centering}>{\columncolor{myblue}}m{0.61cm}<{\centering}>{\columncolor{myblue}}m{0.61cm}<{\centering}>{\columncolor{mygreen}}m{0.61cm}<{\centering}>{\columncolor{mygreen}}m{0.61cm}<{\centering}>{\columncolor{mygreen}}m{0.61cm}<{\centering}>{\columncolor{mygreen}}m{0.61cm}<{\centering}>{\columncolor{mygreen}}m{0.61cm}<{\centering}>{\columncolor{mygreen}}m{0.61cm}<{\centering}>{\columncolor{myblue}}m{0.61cm}<{\centering}>{\columncolor{myblue}}m{0.61cm}<{\centering}>{\columncolor{mygreen}}m{0.61cm}<{\centering}>{\columncolor{mygreen}}m{0.61cm}<{\centering}>{\columncolor{mygreen}}m{0.61cm}<{\centering}>{\columncolor{mygreen}}m{0.61cm}<{\centering}<{\centering}>{\columncolor{mygreen}}m{0.61cm}<{\centering}>{\columncolor{mygreen}}m{0.61cm}<{\centering}}
                \noalign{\global\arrayrulewidth=2pt}
                \toprule
                \multirow{3}*{Method} & \multicolumn{8}{c}{\textbf{Subject-dependent}} & \multicolumn{8}{c}{\textbf{Cross-subject}}\\
                \cmidrule(lr){2-9}\cmidrule(lr){10-17}& \multicolumn{2}{>{\columncolor{myblue}}c}{\textbf{Attention}}& \multicolumn{2}{>{\columncolor{mygreen}}c}{\textbf{Emotion-V}} & \multicolumn{2}{>{\columncolor{mygreen}}c}{\textbf{Emotion-A}}& \multicolumn{2}{>{\columncolor{mygreen}}c}{\textbf{Emotion-D}}&
                \multicolumn{2}{>{\columncolor{myblue}}c}{\textbf{Attention}}& \multicolumn{2}{>{\columncolor{mygreen}}c}{\textbf{Emotion-V}} & \multicolumn{2}{>{\columncolor{mygreen}}c}{\textbf{Emotion-A}}& \multicolumn{2}{>{\columncolor{mygreen}}c}{\textbf{Emotion-D}}\\
                
                \cmidrule(lr){2-3}\cmidrule(lr){4-5}\cmidrule(lr){6-7}\cmidrule(lr){8-9}\cmidrule(lr){10-11}\cmidrule(lr){12-13}\cmidrule(lr){14-15}\cmidrule(lr){16-17}
                &ACC&F1&ACC&F1&ACC&F1&ACC&F1&ACC&F1&ACC&F1&ACC&F1&ACC&F1\\
                \noalign{\global\arrayrulewidth=0.5pt}			
                \midrule
        SVM&78.71&54.82&77.94&43.78&78.56&56.24&64.47&47.19&62.28&43.35&65.78&33.33&61.58&25.92&50.13&48.01\\
        RF&80.55&59.56&78.71&65.45&74.73&65.45&65.54&56.93&56.03&45.21&58.57&41.11&60.09&43.78&44.67&34.33\\
        DT&82.38&57.83&81.92&45.00&75.95&64.85&64.77&55.56&60.12&41.92&58.12&44.44&56.96&32.03&48.42&36.67\\
        EEGNet&85.12&59.64&82.71&63.88&81.66&65.66&71.42&56.84&64.84&45.13&65.18&31.78&65.11&39.36&51.80&22.82\\
        ACRNN&82.53&57.94&81.77&56.21&80.13&47.59&72.01&60.72&63.27&43.69&64.70&46.23&63.25&38.71&51.02&33.78\\
        DGCNN&82.14&47.59&80.92&64.73&83.54&65.45&72.65&55.24&63.91&43.65&63.58&31.86&64.03&39.00&53.70&39.33\\
            \noalign{\global\arrayrulewidth=1pt}
            \bottomrule
            \end{tabular}
            \vspace{-0.4cm}
	}}
\end{table*}
\begin{table}[t]
	\caption{Statistical analysis of emotional labels.}
	\label{statistic}
	\renewcommand{\arraystretch}{1}
	\centering
	{\normalsize{
			\begin{tabular}{m{2.5cm}<{\centering}m{1.5cm}<{\centering}m{1.5cm}<{\centering}m{1.5cm}<{\centering}}
				\noalign{\global\arrayrulewidth=2pt}
				\toprule
				Attention State& Valence&Arousal&Dominance\\
				\noalign{\global\arrayrulewidth=0.5pt}			
				\midrule
				Concentrating&2.70$\pm$0.66&2.30$\pm$1.04&2.86$\pm$1.20\\
				Neutral&2.62$\pm$0.70&1.70$\pm$0.60&2.76$\pm$1.36\\
				Relaxing&3.14$\pm$0.87&1.30$\pm$0.77&2.74$\pm$1.17\\
				\midrule
				$\chi^2$ &35.29&25.10&3.51\\
				\noalign{\global\arrayrulewidth=1pt}
				\bottomrule
			\end{tabular}
	}}
	\vspace{-0.45cm}
\end{table}
\subsection{Quantitative Analysis}
We employ several classical machine learning (SVM~\cite{SVM}, Decision Tree~\cite{DT}, and Random Forest~\cite{RF}), and deep learning (EEGNet~\cite{lawhern2018eegnet}, ACRNN~\cite{tao2020eeg}, and DGCNN~\cite{phan2018dgcnn}) algorithms of the EEG analysis field to perform our dataset's attention and emotion classification tasks. 
Additionally, we conduct multi-label correlation studies, i.e., Chi-squared test and multi-task learning, thereby exploring the association between attention and emotion states.

\subsubsection{\textbf{Attention state classification}}
We conduct attention state classification experiments on our dataset using two common setups: subject-dependent and cross-subject. In the subject-dependent scenario, the first 9 of one subject's trials used for training and the last 3 for testing. In the cross-subject scenario, we apply leave-one-out cross-validation. Green areas of the Table~\ref{Emotion} presents the results of the task. The results indicate that most methods achieve accuracies above 80\% in the subject-dependent setup and above 60\% in the cross-subject setup, confirming the effectiveness of our dataset for attention state classification.

\subsubsection{\textbf{Emotion classification}}
We conduct emotion classification tasks using six classical models, following the same subject-dependent and cross-subject setups as in the attention classification tasks. Blue area of the Table~\ref{Emotion} reports the results of the task. The results show that deep learning algorithms generally outperform machine learning algorithms due to their superior feature extraction capabilities, with each deep learning model excelling in different emotional labels. All methods consistently achieve accuracies well above 33\%, demonstrating the effectiveness of the EEG data and emotional labels for emotion classification.

\begin{figure} [!t]
	\begin{center} 
		
		\subfloat[Subject-dependent] {    
			
			\includegraphics[width=0.46\columnwidth]{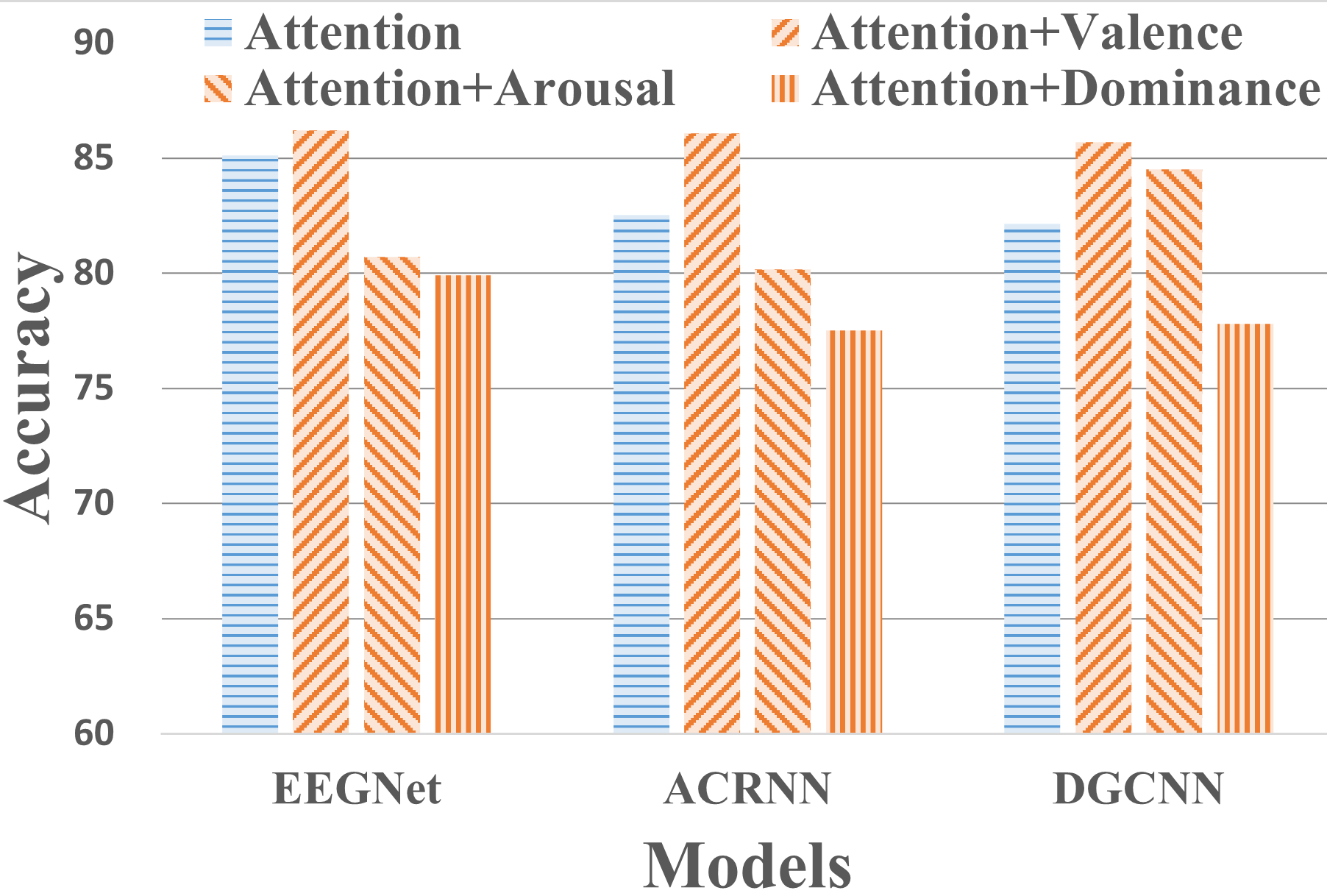}  
		}     
		\hfill
		\subfloat[Cross-subject] {    
			
			\includegraphics[width=0.46\columnwidth]{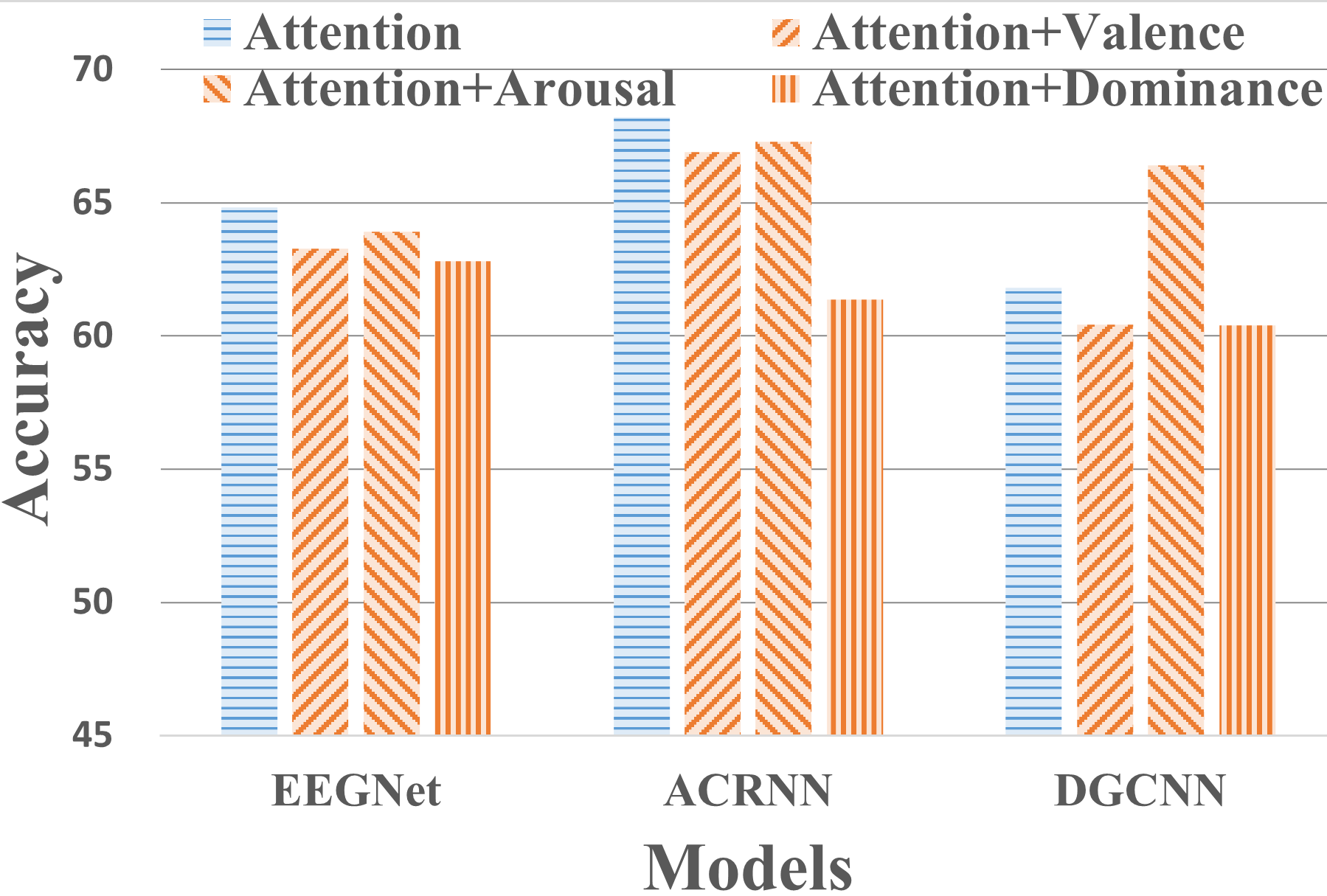}   
		}
	\end{center}
	\vspace{-0.25cm}
	\caption{The comparison of the accuracy of attention state classification in single-task and multi-task scenarios based on three base models with the subject-dependent setting and the cross-subject setting.}
	\label{MT}  
	\vspace{-0.6cm}
\end{figure}
\subsubsection{\textbf{Multi-label correlation study}} 
Firstly, we calculate the average values and standard deviations for the three emotional labels across the three attention states for all subjects, and the Chi-squared test, as shown in Table~\ref{statistic}. The statistical results from the first three rows indicate that attention states have a stronger association with the valence and arousal dimensions, while the association with dominance is weaker. 
This is consistent with the Chi-square test results in the last row~\cite{mchugh2013chi}. Besides, we use Hard Parameter Sharing (HPS)~\cite{zhang2021survey} with three deep learning models to combine attention and emotion classification. 
Figure~~\ref{MT} presents the comparison of attention state classification in single-task and multi-task learning.
The results show that only combining the valence dimension with attention state classification significantly enhances performance across all models in both scenarios. This supports the notion that correct task pairing in multi-task learning boosts performance, while incorrect pairing may cause stagnation or decline~\cite{caruana1997multitask, zamir2018taskonomy}. Research indicates that valence is most closely related to attention states~\cite{hu2016attention}.

\section{Conclusion}
In this paper, we developed an experimental paradigm to gather a multi-label EEG dataset for classifying mental attention states in online learning, based on physiological and psychological principles. We systematically collected EEG data, corresponding labels, and auxiliary information from 20 subjects, totaling over 1,060 minutes of recorded data. Comprehensive qualitative and quantitative analyses confirmed the dataset's high quality and reliability. Our multi-task experiments, incorporating emotional labels, offer deeper insights into the links between attention states and emotions, highlighting the dataset's potential to advance research in this field. Future work will explore the relationship between attention and collected personality traits and expand the dataset by including more subjects and trials.


\section*{Acknowledgment}
This work was supported by the National Natural Science Foundation of China (62192781, 62202367), Project of China Knowledge Centre for Engineering Science and Technology.


\bibliographystyle{IEEEtran}
\bibliography{IEEEabrv,mybibfile}

\end{document}